\documentclass[aps,prl,twocolumn,floats,showpacs,letterpaper,superscriptaddress]{revtex4}
\usepackage{amssymb}
\usepackage{amsbsy}
\usepackage{amsmath}
\usepackage{epsfig}
\usepackage{graphicx}
\usepackage{graphics}
\usepackage{setspace}
\usepackage[utf8]{inputenc}
\usepackage{array}
\usepackage{color}
\usepackage{fontenc}
\usepackage{textcomp}
\usepackage{rotating}
\usepackage{bm}
\usepackage[body={17.5cm,23.0cm},top={2.8cm},left={2.2cm}]{geometry}

\let\s=\sigma

\def\bpm{\begin{pmatrix}}
\def\epm{\end{pmatrix}}
\def\be{\begin{equation}}
\def\ee{\end{equation}}
\def\bea{\begin{eqnarray}}
\def\eea{\end{eqnarray}}
\def\ba{\begin{array}}
\def\ea{\end{array}}

\def\wtd{\widetilde}

\newcommand{\mathsym}[1]{{}}
\newcommand{\unicode}[1]{{}}

\newcommand{\mum}{\mu\mathrm{m}}

\begin{document}

\title{Enhanced Thermoelectric Power in Graphene: Violation of the Mott Relation By Inelastic Scattering }

\author{Fereshte~Ghahari}
\affiliation{Department of Physics, Columbia University, New York, New York 10027, USA}
\author{Hong-Yi~Xie}
\affiliation{Physics and Astronomy Department, Rice University,  Houston, Texas 77005, USA}
\author{Takashi~Taniguchi}
\affiliation{National Institute for Material Science, 1-1 Namiki, Tsukuba, Ibaraki 305-0044, Japan}
\author{Kenji~Watanabe}
\affiliation{National Institute for Material Science, 1-1 Namiki, Tsukuba, Ibaraki 305-0044, Japan}
\author{Matthew~S.~Foster}
\affiliation{Physics and Astronomy Department, Rice University,  Houston, Texas 77005, USA}
\affiliation{Rice Center for Quantum Materials, Rice University, Houston, Texas 77005, USA}
\author{Philip~Kim}
\affiliation{Department of Physics, Columbia University, New York, New York 10027, USA}
\affiliation{Department of Physics, Harvard University, Cambridge, Massachusetts 02138, USA}

\date{\today\\}
\pacs{73.63.-b, 72.80.Vp, 72.15.Jf, 73.23.-b}

\begin{abstract}
We report the enhancement of the thermoelectric power (TEP) in graphene with extremely low disorder. At high temperature we observe that the TEP is substantially larger than the prediction of the Mott relation, approaching to the hydrodynamic limit due to strong inelastic scattering among the charge carriers. However, closer to room temperature the inelastic carrier--optical-phonon scattering becomes more significant and limits the TEP below the hydrodynamic prediction. We support our observation by employing a Boltzmann theory incorporating disorder, electron interactions, and optical phonons.
\end{abstract}

\maketitle

In a diffusive conductor, the electric and thermoelectric transport coefficients can be related by the Mott relation (MR), obtained from the Boltzmann equation in the relaxation-time approximation~\cite{Cutler1969}. The validity of the MR has been tested experimentally for many decades in various materials, such as doped semiconductors~\cite{Ikeda2001},
nanotubes~\cite{Small2003}, nanowires~\cite{Lee2009,Liang2009,Tian2012}, graphene~\cite{Zuev2009,Wei2009}, and topological insulators~\cite{Kim2014}. The technique of tuning the carrier density by electric field effect is particularly useful for examining the MR, since it allows a quantitative comparison between the thermoelectric power (TEP) and electrical conductivity at varying chemical potentials~\cite{Small2003,Lee2009,Liang2009,Tian2012,Zuev2009,Wei2009,Kim2014,Checkelsky2009}.

In the semiclassical regime, the MR is justified as long as the carrier elastic scattering by impurities or quasi-elastic scattering by acoustic phonons dominates the transport~\cite{Hwang2009}. However, the MR should break down when inelastic scattering mechanisms become appreciable at high enough temperatures, when electron-electron (e-e) scattering~\cite{Fritz2008,Muller2008,Foster2009} and carrier--optical-phonon scattering~\cite{Xie2015} become significant. The interaction-dominated thermal and thermoelectric response in graphene was first studied in the context of the hydrodynamic theory of Dirac liquids~\cite{Fritz2008,Muller2008,Foster2009}. At high enough temperature, the enhanced inelastic collisions between charge carriers dramatically accelerate the relaxation towards local thermal equilibrium and yields a hydrodynamic collective behavior. In particular near the charge neutrality point (CNP), the e-e and electron-hole (e-h) scattering rates grow linearly with temperature, and the electron-hole plasma of Dirac fermions develops~\cite{Fritz2008, Muller2008,Foster2009}. In clean graphene, the TEP is simply given by the thermodynamic entropy per carrier charge, which can be substantially higher than the value predicted by the MR, especially, near the CNP~\cite{Fritz2008,Muller2008,Foster2009,Xie2015,Behnia2015}. One also expects that the thermal conductivity violates the Wiedemann-Franz law~\cite{Muller2008,Foster2009}.

The experimental exploration of hydrodynamic transport in graphene began only recently, made possible by the preparation of high-quality graphene samples in hexagonal boron nitride (hBN) encapsulated heterostructures~\cite{Wang2013}. The elastic mean free path $\ell_{\mathrm{el}}$ at low temperature is limited merely by the sample size, typically on the order of 10~$\mu$m. As temperature grows, $\ell_{\mathrm{el}}$ is gradually suppressed to $\sim$1~$\mu$m at room temperature, which is attributed to the thermally enhanced quasi-elastic scattering off acoustic phonons~\cite{Hwang2007,Efetov2010}.
On the other hand, the inelastic e-e scattering length $\ell_{\mathrm{ee}}$ is suppressed more rapidly. At sub-Kelvin temperatures, $\ell_{\mathrm{ee}}$ is experimentally estimated to be $\ell_{\mathrm{ee}} \sim 10$-$100 \,\mum$ ~\cite{Engels2014}. For temperatures above $100$~K, $\ell_{\mathrm{ee}}\lesssim 0.5$~$\mu$m according to theoretical calculations~\cite{Schutt2011,Li2013,Principi2015}. Therefore, the hydrodynamic condition $\ell_{\mathrm{ee}} \ll \ell_{\mathrm{el}}$ is potentially achievable by elevating temperature.

Two recent experiments have indeed observed the fingerprint of hydrodynamic transport in graphene. Crossno \emph{et al.} reported the strongly enhanced thermal conductivity in very clean samples near the CNP compared to the Wiedemann-Franz result~\cite{Crossno2015}, in quantitative agreement with the hydrodynamic theory prediction. Bandurin \emph{et al.} indirectly investigated the hydrodynamic viscosity of electrons away from the CNP by nonlocal charge transport measurement and suggested that submicrometer-size electron vortices should develop at high temperature~\cite{Bandurin2015}. Since the TEP can be related to transport entropy by charged carriers~\cite{Behnia2015}, careful measurement of the TEP in the clean limit samples can provide an important experimental probe for the hydrodynamic flow of Dirac fluid.

\begin{figure}[t]
\centering
\includegraphics[width=1.0\linewidth]{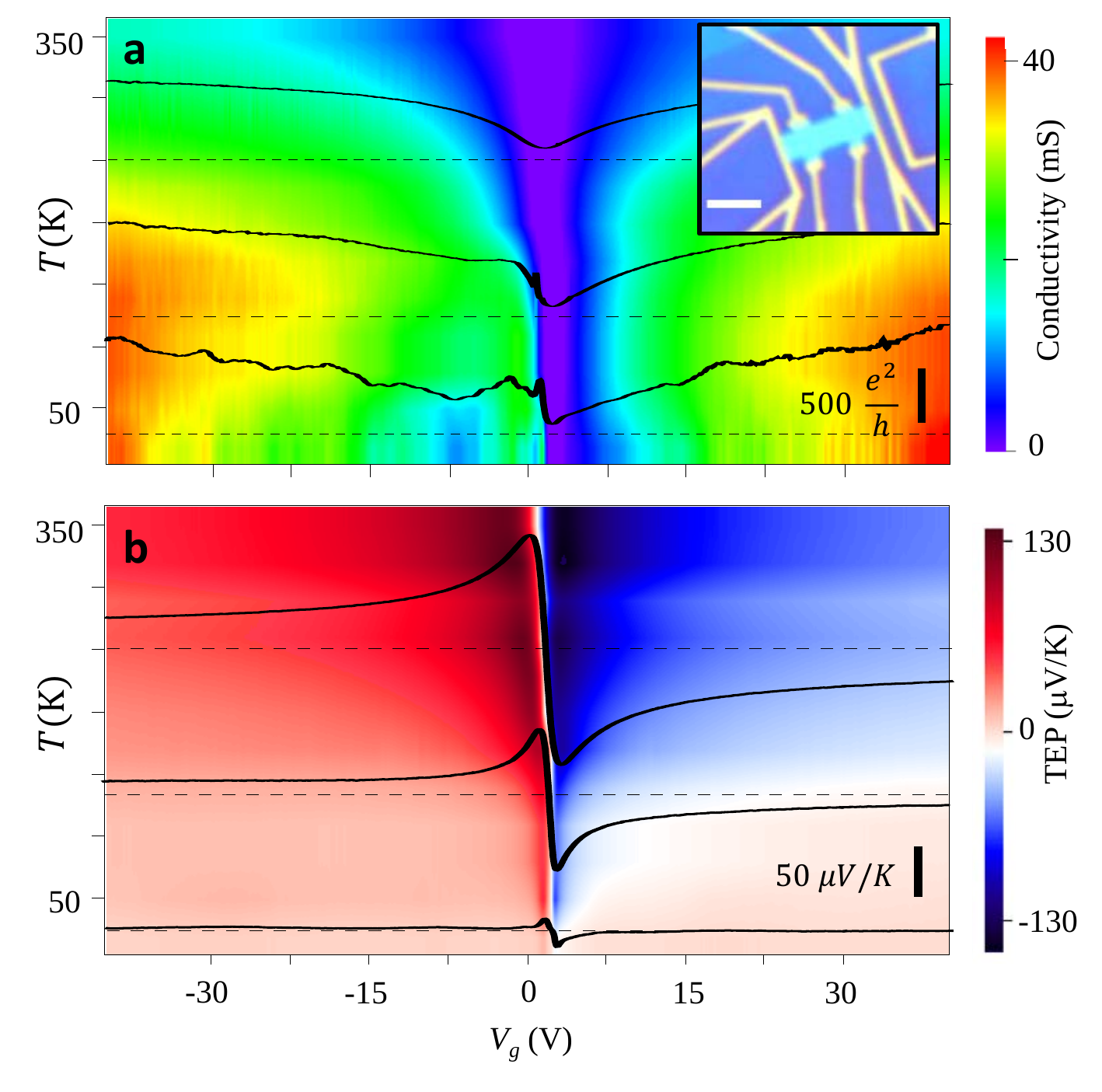}
\caption{(a) the measured conductivity $\s$ (a) and (b) TEP $S$ as functions of temperature $T$ and gate voltage $V_\mathrm{g}$ in the sample with lowest disorder in this experiment (sample $\mathrm{D}_3$). At $T=20$~K, 130~K, and 250~K (horizontal dash cuts), $\s(V_\mathrm{g})$ and $S(V_\mathrm{g})$ (solid curves) are shown in the overlaid graphs where the temperature cuts indicate $\s$ or $S =0$. The upper inset shows a typical device image where the scale bar corresponds to 2 $\mu$m.} \label{5.1}
\end{figure}

In this Letter, we probe the signature of inelastic scattering processes in the thermoelectric response measured in high-mobility graphene samples fabricated on a hBN substrate. We measure the TEP as a function of gate voltage for a broad range of temperatures  (25$<T<$350~K) in samples of varying quality (quenched disorder). We observe that at higher mobilities the TEP is significantly enhanced above the MR estimation, but persistently below the hydrodynamic prediction.
We explain this anomalous feature of the TEP using Boltzmann transport theory including the scattering mechanisms due to disorder, electron-electron interactions, and intrinsic optical phonons of graphene. We suggest that at high temperature ($T \gtrsim 200$~K) the optical phonons become the major obstacle to achieving the hydrodynamic transport in graphene.

The graphene devices used in our experiment were prepared by the method introduced in Refs.~\cite{Wang2013, Dean2010}, where the hBN is employed as a substrate or encapsulation. A typical TEP device image is shown in the inset of Fig.~1(a).
The TEP is measured by $S(V_\mathrm{g},T)=-\delta V/\delta T$, where $\delta T$ is a controlled temperature gradient applied to the sample and $\delta V$ is the thermally induced voltage across the sample. A gate voltage $V_\mathrm{g}$ is applied to the silicon substrate to tune the carrier density $n$ in the graphene channel, where the gate capacitance coupling $C_\mathrm{g}=110$~aF/$\mum^2$ is obtained from the Hall measurement. The source and drain contacts for the graphene channel
also serve as resistance thermometers in a four-terminal geometry. The four-terminal conductivity $\sigma(V_\mathrm{g},T)$ is obtained in the Hall bar geometry. Technical details can be found in Ref.~\cite{Zuev2009}. We measure three samples $\mathrm{D}_{1,2,3}$, with decreasing disorder labeled $1$ to $3$, in the temperature range $25<T<300$~K. We estimate the mobility of the samples $\mathrm{D}_{1,2,3}$ as $\sim 2.5$, $5$, and $10$~m$^2$/Vsec, respectively, at $T \sim 100$ K.

In Fig.~\ref{5.1}(a) we show the measured conductivity $\sigma(V_\mathrm{g},T)$ in sample $\mathrm{D}_3$. At fixed temperature, $\sigma(V_\mathrm{g})$ exhibits minima at $V_\mathrm{g}=V_\mathrm{D} \approx 0$ corresponding to the CNP.
At large gate voltage, $\sigma(T)$ reaches maxima at $T\approx 100$~K, where we estimate $\ell_{\mathrm{el}}\sim$2~$\mu$m
(about the sample size). Below $100$~K, visible mesoscopic fluctuations start developing and become stronger with lowering temperature, presumably due to the scattering of carriers at the sample boundary. In Fig.~\ref{5.1}(b) we show the measured TEP $S(V_\mathrm{g},T)$ in sample $\mathrm{D}_3$. $S(V_\mathrm{g})$ changes sign across the CNP as the carrier type changes from electrons to holes. Moreover, $|S(V_\mathrm{g})|$ exhibits almost symmetric maxima about the CNP. The peak values are $\sim$100~$\mu$V/K at room temperature, somewhat higher than the values observed in more disordered samples in previous studies~\cite{Zuev2009,Wei2009}.

We first introduce the MR that is used to estimate the TEP in our analysis. In the present high-quality samples, the charge density fluctuation can be suppressed as much as $\delta n\sim 10^{10}$~cm$^{-2}$, leading to the chemical potential fluctuation much below the experimental temperatures $\delta \mu\lesssim 10$~meV~\cite{Crossno2015}. Therefore, we exploit the general MR~\cite{Cutler1969,Ashcroft&Mermin},
\be \label{Generalizedmott}
	S_{\mathrm{Mott}}
	=
	-\frac{1}{|e| T}
	\frac{
	\int_{-\infty}^{\infty}(\epsilon-\mu)\sigma(\epsilon)\frac{\partial f(\epsilon)}{\partial\epsilon}d\epsilon
	}{
	\int_{-\infty}^{\infty}\sigma(\epsilon)\frac{\partial f(\epsilon)}{\partial\epsilon}d\epsilon
	},
\ee
where $e$ is the electron charge,
$f(\epsilon)=1/[e^{(\epsilon-\mu)/{k_{\mathrm{B}} T}}+1]$ the equilibrium Fermi-Dirac distribution function, and $\sigma(\epsilon)$ the energy-dependent conductivity kernel. Analyzing our data, we replace $\sigma(\epsilon)$ with the measured conductivity at the gate voltage determined by
$
	\epsilon=E_{\mathrm{F}} = \hbar v_F\sqrt{\pi C_\mathrm{g} \delta V_\mathrm{g} / |e|}
$,
where $v_F=10^6$m/s is the Fermi velocity of graphene, $\delta V_\mathrm{g}=V_\mathrm{g}-V_\mathrm{D}$, and $E_{\mathrm{F}}$ the Fermi energy. It has been demonstrated that this method provides a unified way to examine the MR in graphene at any doping and temperature, including the vicinity of the CNP~\cite{Wang2011}. We note that, in the degenerate regime $k_{\mathrm{B}} T \ll \mu$, the MR~(\ref{Generalizedmott}) reduces to the more familiar differential form
$
	S_{\mathrm{Mott}}
	=
	-
	\frac{\pi^{2}k_{\mathrm{B}}^{2}T}{3|e|}
	\frac{1}{\sigma} \frac{ \mathrm{d} \sigma }{ \mathrm{d} V_{\mathrm{g}} }
	\frac{ \mathrm{d} V_{\mathrm{g}} }{ \mathrm{d} E_{\mathrm{F}} }
$,
that has been verified extensively~\cite{Zuev2009,Wei2009,Checkelsky2009}.

\begin{figure}[t]
\centering
\includegraphics[width=1.0\linewidth]{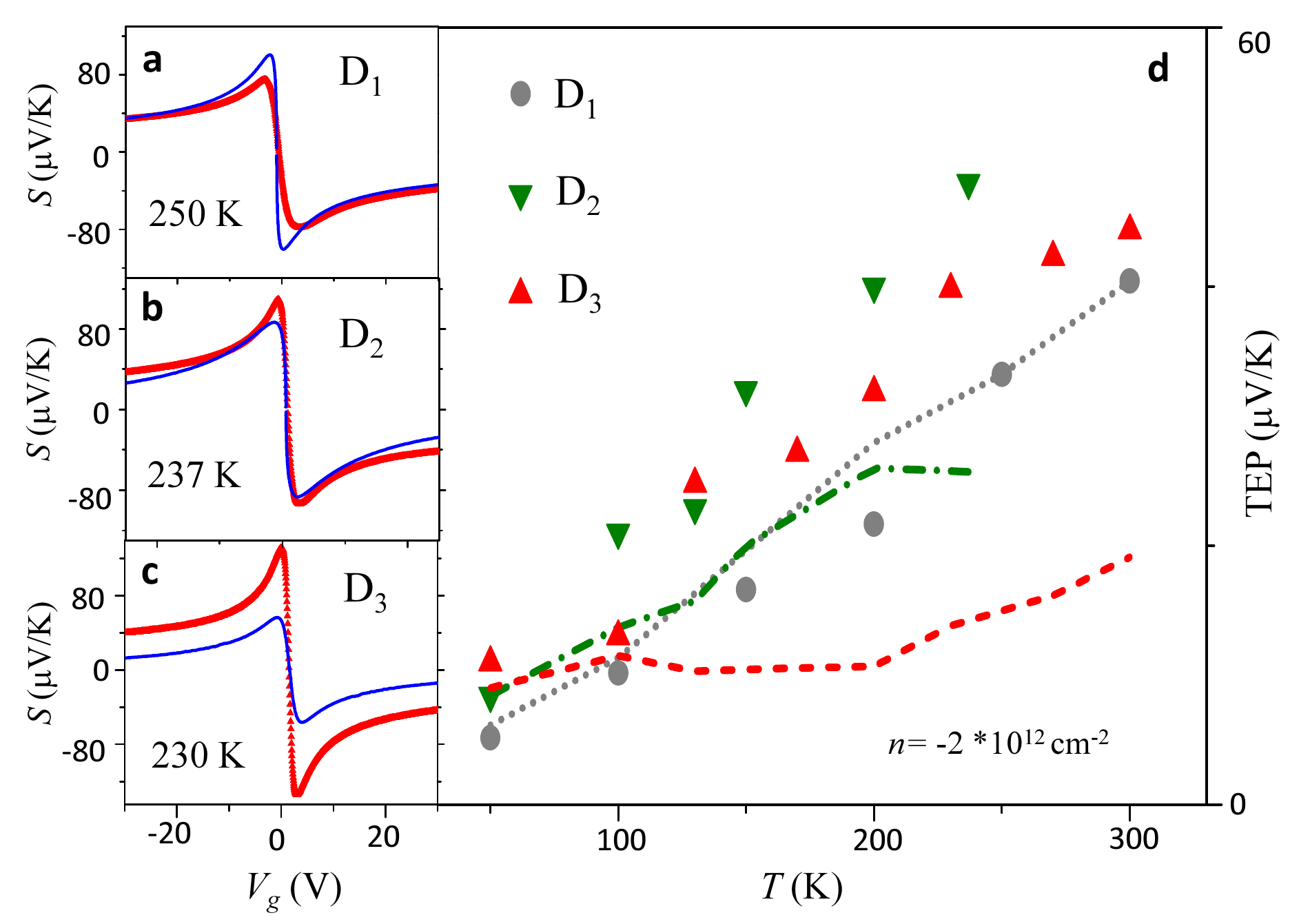}
\caption{Comparison between the measured TEP and the MR estimation Eq.~(\ref{Generalizedmott}) in samples $\mathrm{D}_{1,2,3}$. (a)-(c) The measured (red symbols) and MR estimated (blue symbols) TEP as functions of $V_\mathrm{g}$ at fixed high temperatures. (a), (b), and (c) are for samples D$_{1,2,3}$ at $T=250~\mathrm{K}$, $237~\mathrm{K}$, and $250~\mathrm{K}$, respectively. (d) The measured (symbols) and MR estimated (lines) TEP as functions of $T$ at a fixed carrier density of $ n = 2 \times 10^{12}$cm$^{-2}$. The dotted, dash-dotted and dashed lines indicates the MR estimation for samples $\mathrm{D}_{1,2,3}$, respectively.
} \label{5.2}
\end{figure}

In Figs.~\ref{5.2}(a--c) we compare the measured TEP to the MR estimation [Eq.~(\ref{Generalizedmott})] for samples $\mathrm{D}_{1,2,3}$. A representative high-temperature limit $T\gtrsim 230$~K is chosen for all the samples to make a contrast. The MR value $S_\mathrm{Mott}$ exhibits a notable deviation trend from the measured TEP $S$, depending on the degree of disorder. For the lowest quality sample $\mathrm{D}_{1}$, $S$ coincides qualitatively well with $S_\mathrm{Mott}$.
Only a small deviation near the CNP $S/S_\mathrm{Mott}\lesssim 1$ is noticeable, similar to the observation in the previous work~\cite{Wang2011}. This discrepancy is possibly due to the overestimation of the temperature effect in $\sigma(\epsilon)$ that is simply replaced by the measured conductivity~\cite{Wang2011}. However, as sample quality improves further, a new trend emerges. For the medium quality sample D$_2$, where we estimate $\ell_{\mathrm{ee}} \lesssim \ell_{\mathrm{el}}$, we find that $S/S_\mathrm{Mott} \gtrsim 1$ but the discrepancy is relatively small ($<20$\%). For the highest quality sample D$_3$, where $\ell_\mathrm{ee}\ll\ell_\mathrm{el}$, we observe that $S/ S_{\mathrm{Mott}} \approx 2$. This strong enhancement indicates the violation of the MR in low-disorder samples at high temperature.

As shown in Fig.~\ref{5.2}(d), we further investigate the TEP enhancement by looking into the temperature dependence at a fixed density. The most disordered sample D$_1$ exhibits no appreciable deviation from the MR prediction in the entire temperature range $50<T<300$~K. For higher quality samples the deviation becomes more significant with elevating temperature. For sample D$_3$, the enhancement factor $S/S_{\mathrm{Mott}} \gtrsim 2$ when $T>100$~K. We also notice that $S(T)$ tends to increase linearly with $T$, which suggests that the phonon-drag effect in our devices is not significant, unlike the observation in GaAs heterostructures~\cite{Basu1988,Nicholas1985,Cantrell1986}.

\begin{figure}[t]
\centering
\includegraphics[width=1.0\linewidth,]{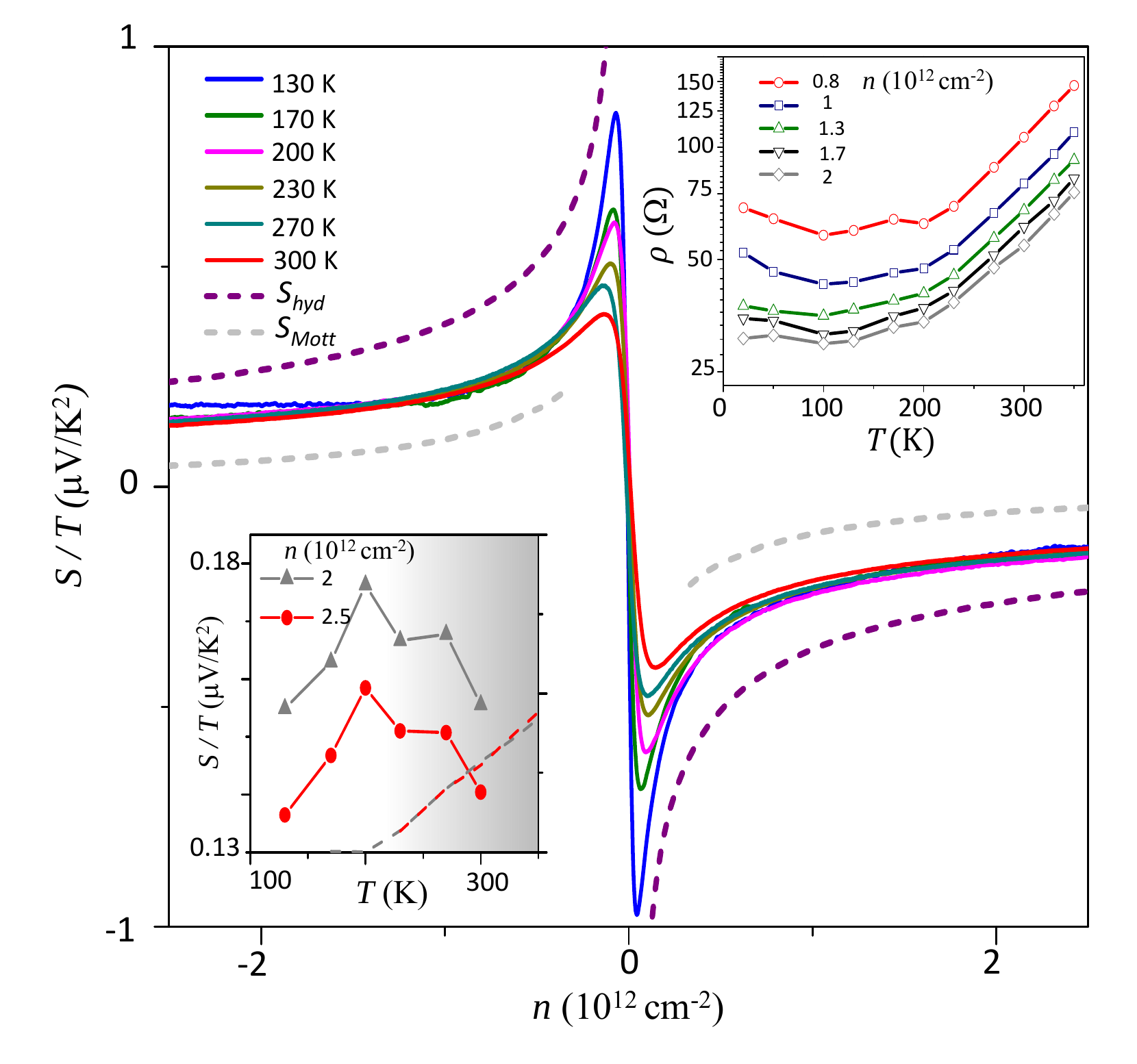}
\caption{The measured $S/T$ as a function of density $n$ for various temperatures in sample D$_{3}$. The gray and purple dash curves show $S_\mathrm{Mott}/T$ [Eq.~(\ref{Generalizedmott})] and $S_\mathrm{hyd}/T$ [Eq.~(\ref{hydroclean})], respectively. The upper inset shows the resistivity $\rho(T)$ for various densities. The lower inset shows the measured $S/T$ as a function of $T$ at two densities $n = 2 \times 10^{12}\,\mathrm{cm}^{-2}$ and $2.5 \times 10^{12}$cm$^{-2}$.}\label{5.3}
\end{figure}

The violation of the MR expressed in Eq.~(\ref{Generalizedmott}) is potentially attributed to the predominance of inelastic scattering processes at high temperature in clean samples. For instance, in the ideal hydrodynamic regime, where the e-e interaction is the only scattering mechanism, the TEP $S_\mathrm{hyd}$ is theoretically predicted to be the thermodynamic entropy per charge of the carriers~\cite{Fritz2008,Muller2008,Foster2009,Xie2015}. In the degenerate limit $k_BT\ll\mu$, approximating the entropy density as that of the ideal Fermi gas with linear dispersion gives~\cite{Foster2009, Muller2008}
\begin{equation}
	S_\mathrm{hyd}\approx\frac{2\pi^2}{3}\frac{k_\mathrm{B}^2 T}{|e| \mu}.
	\label{hydroclean}
\end{equation}
We note that this hydrodynamic TEP can be substantially larger than that in a diffusive conductor (see Fig.~\ref{5.3} and discussion in Ref.~\onlinecite{Xie2015}).

In Fig.~\ref{5.3}, we contrast the measured TEP $S$ to the Mott limit $S_{\mathrm{Mott}}$ [Eq.~(\ref{Generalizedmott})] and
the hydrodynamic limit $S_\mathrm{hyd}$ [Eq.~(\ref{hydroclean})]. The ratio $S/T$ is plotted as a function of the carrier density $n$ at various temperatures measured in sample D$_3$. We observe that, besides the enhancement above $S_\mathrm{Mott}$, $S$ is also significant below $S_\mathrm{hyd}$ for densities $n < 2.5 \times 10^{12}\,\mathrm{cm}^{-2}$
over the temperature range $130<T<300$~K. An examination of the temperature dependence of $S/T$ at fixed density reveals more features. The lower inset of Fig.~\ref{5.3} shows $S/T$ in the temperature range $100<T<300$~K measured at two fixed densities $n =2 \times 10^{12}$cm$^{-2}$ and $2.5 \times 10^{12}$cm$^{-2}$. We find that for both densities the $S/T$ ratio exhibits maxima around $T = T^\ast\approx 200$ K: $S/T$ grows towards the hydrodynamic limit for $T<T^\ast$, possibly because of the suppression of $\ell_{\mathrm{ee}}$, but turns to decay when $T > T^\ast$. The nonmonotonicity of $S/T$ in temperature strongly suggests that other inelastic scattering processes besides the e-e interaction become nonnegligible above $T^\ast$.
We note that the intercarrier collisions responsible for the hydrodynamic linear response are special, because there is no preferred rest frame for the Dirac fluid. This is not typically true for other inelastic scattering mechanisms.

The optical phonons may serve as an important source of total momentum relaxation below room temperature. This can be further substantiated by the experimental observation of the superlinear temperature dependence of the resistivity $\rho(T)$ at $T \gtrsim T^\ast$ shown in the upper inset of Fig.~\ref{5.3}. Recent ab-initio calculation suggests that the high-temperature superlinearity of $\rho(T)$ can be attributed to carrier scattering off optical phonons of graphene~\cite{Park2014}. Although the energy scale of the optical phonon ($\sim$0.15~eV) is well above the room temperature, strong electron--optical-phonon coupling can provide a substantial activated effect in $\rho(T)$ above $T^\ast$.~\cite{Park2014} This superlinearity of $\rho(T)$ has been observed also in other experiments~\cite{Efetov2010,Morozov2008,Chen2008}.

We employ a theoretical model based on the Boltzmann equation (for the carrier distribution function $f$)
characterized by the collision integral~\cite{Xie2015},
\begin{equation}  \label{bolt-coll}
	\mathfrak{St}[f]
	=
	\mathfrak{St}_{\mathrm{el}}[f]
	+
	\mathfrak{St}_{\mathrm{int}}[f]
	+
	\mathfrak{St}_{\mathrm{oph}}[f].
\end{equation}
Here $\mathfrak{St}_{\mathrm{el}}[f]$ describes the elastic scattering off short-ranged impurities, characterized by an effective disorder strength $\wtd{g}$, and the scattering off screened Coulomb impurities, controlled by the graphene fine structure constant $\alpha_{G}$ and the impurity concentration $n_{\mathrm{imp}}$. $\mathfrak{St}_{\mathrm{int}}[f]$ represents the Coulomb collisions incorporating the processes
``$\mathrm{ e + e  \leftrightarrow e  + e }$'' (intraband, Channel A)
and
``$\mathrm{ e+h \leftrightarrow e+h}$'' (interband, Channel B),
including temperature and density-dependent screening effects.
Here ``e'' (``h'') denotes a conduction band electron (valence band hole).
Finally, $\mathfrak{St}_{\mathrm{oph}}[f]$ describes carrier--optical-phonon scattering processes
``$\mathrm{ph + e \, (h) \leftrightarrow e \, (h) }$''
and
``$ \mathrm{e + h \leftrightarrow ph}$'', where ``ph'' denotes an optical phonon~\cite{footnote1}.
We take $\alpha_{G}=e^2/\kappa\hbar v_F\approx 0.6$ using the dielectric constant of the hBN encapsulation $\kappa \approx 3.8$. Solving the linearized Boltzmann equation we obtain the conductivity $\s(n,T)$ and the TEP $S(n,T)$. Theoretical details will appear elsewhere~\cite{Xie2015}.

\begin{figure}[b]
\centering
\includegraphics[width=1.0\linewidth]{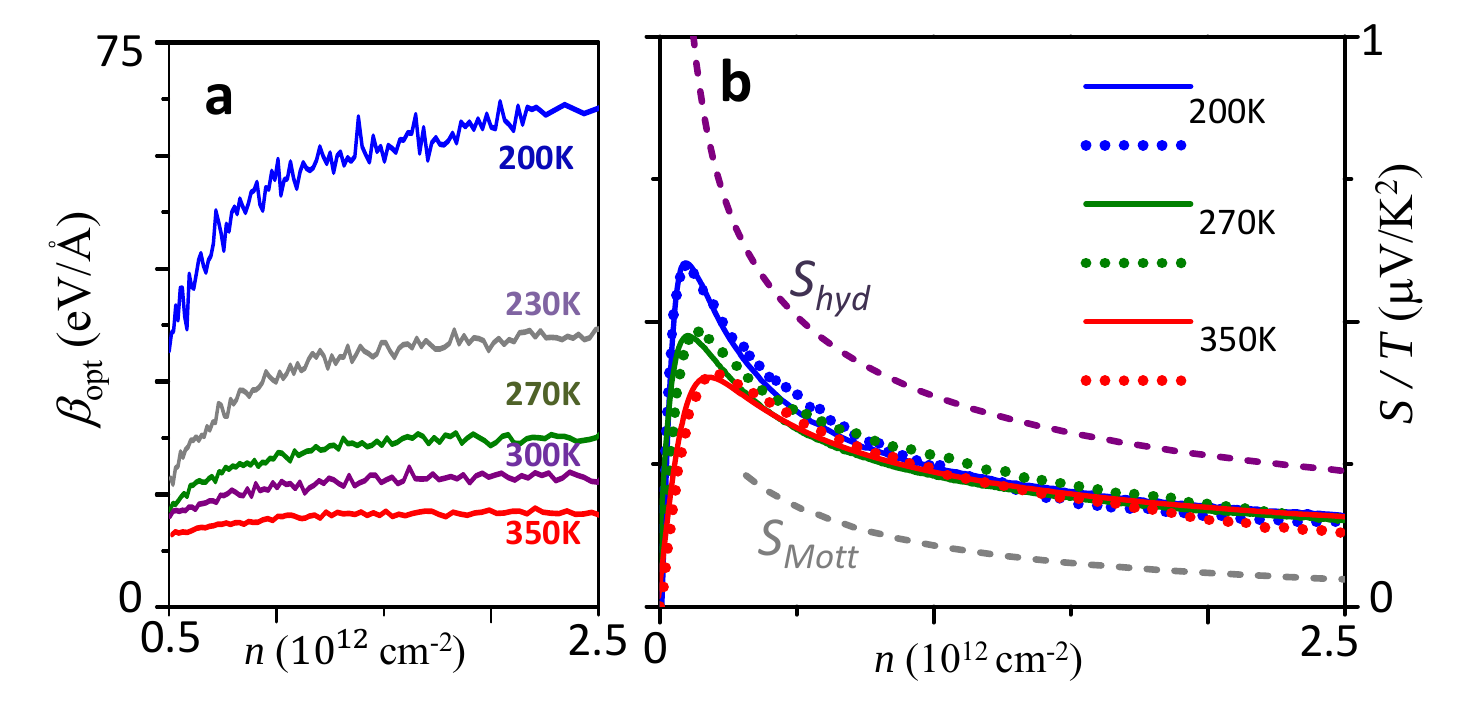}
\caption{(a) Electron-optical phonon coupling $\beta^{2}_{\mathrm{opt}} (n, T)$ extracted form the conductivity data shown in Fig.~1(a). (b) $S/T$ as a function of density $n$ at three temperatures $T$=50~K, 170~K and 300~K (solid curves). The gray and purple dash curves show $S_\mathrm{Mott}$ and $S_\mathrm{hyd}$, respectively. The dotted curves represent the prediction of the Boltzmann model Eq.~(\ref{bolt-coll}).} \label{5.5}
\end{figure}

In our analysis, for simplicity we consider a single Einstein optical phonon mode that remains in equilibrium. We have in mind the $A^{\prime}_{1}$ mode \cite{Sohier2014,Manes2007,Basko2008}, which corresponds to the ``kekule'' vibration of the honeycomb lattice and scatters electrons between valleys. This is suggested to be the most relevant branch for transport at low temperature, possessing the lowest excitation energy $T_{A^\prime} = \hbar\omega/k_B \approx 1740$~K and the largest electron-phonon coupling strength $\beta_{A^{'}}$~\cite{Sohier2014}. In practice, we fit $\beta_{A^{'}}$ from the conductivity data [see Fig.~\ref{5.5}]. To achieve the best \emph{quantitative} agreement with the data, we set $T_{A^\prime} = \hbar\omega/k_B \approx 2200$~K. The reason for this enhancement might be that the $A_{1}'$ phonons are more rigid due to the encapsulation, or that higher-frequency optical-phonon branches are also involved.

We analyze the sample D$_3$ by first fixing the model parameters from the measured conductivity. The short-ranged impurity strength $\wtd{g}$ and Coulomb impurity density $n_\mathrm{imp}$ are determined by the conductivity at low temperature and high doping, where the effects of electron-electron interaction and optical phonons are not significant. The electron-optical phonon coupling parameter $\beta_A^{\prime}$ is estimated by fitting the conductivity data in the high-temperature regime ($T>170$~K). In Fig.~4(a) $\beta_{A^{\prime}}(n,T)$ is shown as a function of density at various temperatures. We observe that $\beta_{A^{\prime}}$ is almost density-independent for $n>1\times 10^{12}$cm$^{-2}$, due to strong Thomas-Fermi screening in this regime~\cite{Xie2015}, but significantly increases with decreasing $T$, likely due to the Coulomb renormalization of the coupling strength~\cite{Xie2015,BaskoAleiner2008,Basko2008}.

Using these model parameters we calculate the TEP and compare the result to the experimental data in Fig.~4(b). The theoretical result fits the data in the whole range of densities including the non-degenerate limit near the CNP. This quantitative agreement supports our conjecture that the inelastic scattering by optical phonons suppresses the TEP from the hydrodynamic limit at relatively low temperatures $T/T_\mathrm{opt} \sim 0.1$. We argue that, in a very clean sample, the TEP becomes sensitive to the scattering off optical phonons as soon as the carrier--optical-phonon scattering length becomes comparable to $\ell_{\mathrm{el}}$, despite the fact that $T\ll T_\mathrm{oph}$~\cite{Xie2015}.

In summary, we observe that in clean graphene samples the TEP at high temperature is enhanced substantially beyond the MR. The observation can be explained by the inclusion of the prevailing inelastic scattering due to both Coulomb interaction among charge carriers and electron--optical-phonon coupling.

H.-Y.~X. and M.~S.~F. thank Jesse Crossno, Kin Chung Fong, and Markus M\"uller for stimulating discussions.  This major experimental work is supported by DOE (No. DE-SC0012260). P.K. acknowledges a partial support from the Nano Material Technology Development Program through the National Research Foundation of Korea (NRF) funded by the Ministry of Science, ICT and Future Planning (2012M3A7B4049966). H.-Y.~X. and M.~S.~F. were supported by the Welch Foundation under Grant No.~C-1809 and by an Alfred P. Sloan Research Fellowship (No.~BR2014-035).


\begin{thebibliography}{99}
\bibitem{footnote1} We include the imbalance collision process ``$ \mathrm{e + h \leftrightarrow ph}$,'' which breaks the separate particle- and hole-number
conservation. The process is suppressed at finite doping but becomes important around the CNP.
\end{thebibliography}
\end{document}